\documentclass[twocolumn,twocolappendix,tighten,times,astrosymb]{aastex631}

\usepackage{amsmath}
\usepackage{soul}
\usepackage{microtype}
\usepackage[english]{babel}
\usepackage{cancel}
\usepackage{soul}
\hypersetup{linkcolor=blue,citecolor=blue,filecolor=blue,urlcolor=blue}

\newcommand{\be}{\begin{equation}}
\newcommand{\ee}{\end{equation}}

\newcommand{\lcrg}{(L/5r_{\rm g})}

\newcommand{\sigmat}{\sigma_{\rm T}}

\long\def\exclude#1{}

\shorttitle{TeV neutrinos from NGC 1068}
\shortauthors{}
\begin{document}

\title{\large TeV neutrinos and hard X-rays from relativistic reconnection in the corona of NGC 1068}

\correspondingauthor{damiano.fiorillo@nbi.ku.dk }
\author[0000-0003-4927-9850]{Damiano F. G. Fiorillo}
\affiliation{Niels Bohr International Academy, Niels Bohr Institute, University of Copenhagen, 2100 Copenhagen, Denmark}
\author[0000-0001-6640-0179]{Maria Petropoulou}
\affiliation{Department of Physics, National and Kapodistrian University of Athens, University Campus Zografos, GR 15784, Athens, Greece }
\affiliation{Institute of Accelerating Systems \& Applications, University Campus Zografos, Athens, Greece}
\author[0000-0001-8822-8031]{Luca Comisso}
\affiliation{Department of Astronomy and Columbia Astrophysics 
Laboratory, Columbia University, New York, NY 10027, USA}
\author[0000-0003-0543-0467]{Enrico Peretti}
\affiliation{Niels Bohr International Academy, Niels Bohr Institute, University of Copenhagen, 2100 Copenhagen, Denmark}
\affiliation{Université Paris Cité, CNRS, Astroparticule et Cosmologie, 10 Rue Alice Domon et Léonie Duquet, F-75013 Paris, France}
\author[0000-0002-1227-2754]{Lorenzo Sironi}
\affiliation{Department of Astronomy and Columbia Astrophysics 
Laboratory, Columbia University, New York, NY 10027, USA}
\affiliation{Center for Computational Astrophysics, Flatiron Institute, 162 5th Avenue, New York, NY 10010, USA}

\begin{abstract}
The recent discovery of astrophysical neutrinos from the Seyfert galaxy NGC 1068 suggests the presence of non-thermal protons within a compact ``coronal'' region close to the central black hole. The acceleration mechanism of these non-thermal protons remains elusive. We show that a large-scale magnetic reconnection layer, of the order of a few gravitational radii, may provide such a mechanism. In such a scenario, rough energy equipartition between magnetic fields, X-ray photons, and non-thermal protons is established in the reconnection region. Motivated by recent three-dimensional particle-in-cell  simulations of relativistic reconnection, we assume that the spectrum of accelerated protons is a broken power law, with the break energy being constrained by energy conservation (i.e., the energy density of accelerated protons is at most comparable to the magnetic energy density). The proton spectrum is $dn_p/dE_p\propto E_p^{-1}$ below the break, and $dn_p/dE_p\propto E_p^{-s}$ above the break, with IceCube neutrino observations suggesting $s \simeq 3$.
Protons above the break lose most of their energy within the reconnection layer via photohadronic collisions with the coronal X-rays, producing a neutrino signal in good agreement with the recent observations. Gamma-rays injected in photohadronic collisions are cascaded to lower energies, sustaining the population of electron-positron pairs that makes the corona moderately Compton thick. 
\end{abstract}


\section{Introduction}

Active galactic nuclei (AGN) are among the most energetic sources in the Universe. They are powered by accreting supermassive black holes (SMBHs), and  
their main emission features \citep[see, e.g.,][]{Marconi_2004_SED,Ghisellini_2013,Padovani:2017zpf} can be summarized as: (i) an optical-ultraviolet thermal component, often referred to as the ``big blue bump'', which is produced by the accretion disk; (ii) a second thermal component radiated by the dusty torus in the infrared; (iii) an X-ray power law extending up to $100$~keV, which is believed to be associated with the AGN ``coronal'' region.

 The coronal region of AGN has been proposed as a production site of multi-messenger emission 
 since the late seventies \citep{Berezinsky_1977,Silberberg_1979,Eichler_1979}. Very recently, the IceCube collaboration has reported compelling evidence of a high-energy neutrino flux from NGC~1068 \citep[][]{Aartsen_Icecube_2020,IceCube-NGC1068}, a nearby Seyfert II galaxy located at a distance of $\sim$10.1 Mpc \citep[][and Padovani et al., in prep.]{Tully_2008}. 
This neutrino flux has been measured with a significance of 4.2$\sigma$ in the energy band 1.5--15 TeV, showing a soft neutrino spectrum $d\Phi_\nu/dE_\nu\propto E_\nu^{-3.2}$.
A comparable TeV gamma-ray flux, which would be expected for an optically thin source, is not observed, with stringent upper limits set by the MAGIC telescope~\citep{MAGIC-UL-NGC1068}. 
This suggests that the neutrino source must be located close to the AGN innermost regions, where the dense optical and X-ray radiation would reprocess gamma rays to lower energies~\citep{Berezinsky_1981}. Thus, the GeV gamma rays observed by Fermi-LAT~\citep{Abdo2010} are hard to explain within the coronal model, and are generally attributed to a different region, e.g., the weak jet observed in the core of the galaxy \citep{Lenain2010}, the circumnuclear starburst \citep{Yoast-Hull2013,Ambrosone:2021aaw,Eichmann_2022}, a large scale AGN-driven outflow \citep{Lamastra_2016}, a combination of successful and failed line-driven wind \citep{Inoue_S_2022}, or an ultra-fast outflow \citep{Peretti_2023}.

Neutrino emission from AGN cores directly points to the presence of a relativistic hadronic population, but the specific particle acceleration mechanism at work remains an open question.
\citet{Inoue_2020} and \cite{Murase:2019vdl} provided the first phenomenological modeling of NGC~1068, prescribing that protons are energized via diffusive shock acceleration 
(DSA) or stochastic acceleration, respectively.
Both models highlighted the importance of Bethe-Heitler (BH) interactions on optical photons as the main mechanism limiting proton acceleration. \citet{Kheirandish_2021} discussed magnetic reconnection as a potential mechanism for proton acceleration in a weakly magnetized corona. Assuming a proton power-law spectrum with slope $-2$, they computed the neutrino emission from $pp$ and $p\gamma$ inelastic collisions, and applied their analysis to NGC 1068. 
Recently, \citet{Murase_2022ApJL} provided model-independent multi-messenger constraints on the acceleration processes, as well as on the production mechanism of high-energy neutrinos in AGN coronae. This study highlighted the importance of the electromagnetic cascade for  reprocessing the GeV-TeV gamma rays into the MeV band.

The aim of this Letter is to explore the role of reconnection in magnetically-dominated plasmas for
powering the X-ray and neutrino emission in AGN coronae. In this scenario, rough energy equipartition between magnetic fields, X-ray radiation, and relativistic protons is established in the reconnection region. 
Motivated by recent three-dimensional particle-in-cell (PIC) simulations of relativistic reconnection~\citep{Zhang:2021akj,Zhang:2023lvw,Chernoglazov:2023ksi}, we assume that the spectrum of accelerated protons is a broken power law, with the break energy $E_{p, \rm br}$ being constrained by energy conservation (i.e., the energy density of accelerated protons is at most comparable to the magnetic energy density)\footnote{As shown below, the break energy is well below the maximum proton energy achievable in reconnection.}. The proton spectrum is hard below the break (slope around $-1$), whereas it is softer above the break, which could explain the soft neutrino spectrum of NGC 1068. Our model has only two free parameters, namely the size of the reconnection region, and the number density of non-thermal protons in the coronal region, which directly determines $E_{p, \rm br}$. All other parameters can be either inferred by observations or benchmarked with PIC simulations. 

In this Letter we demonstrate that the neutrino emission expected from relativistic reconnection in the coronal region of NGC 1068 is consistent with  IceCube observations, and we show that the peak neutrino luminosity scales quadratically with the X-ray luminosity and inversely proportional to the black hole mass. For large X-ray luminosities, protons cool so fast that they lose most of their energy inside the reconnection layer, which effectively becomes a calorimeter: in this regime, the scaling saturates into a linear proportionality with the X-ray luminosity, independently of the black hole mass and layer size. We find that Bethe-Heitler and photohadronic interactions initiate an electromagnetic cascade that ultimately establishes and maintains a population of electron-positron pairs, which naturally makes the corona moderately optically thick for the Comptonization of low-energy photons up to the hard X-ray band.

\section{Properties of the reconnection layer}\label{sec:global_properties}

In this section we identify the main properties of the reconnection layer. Assuming that the hard X-ray flux comes from Comptonization within the reconnection layer, we can infer the electron number density, and also provide an estimate for the magnetic field strength in the reconnection layer.

We envision the formation of a reconnection layer in the vicinity of a black hole with mass $M=10^7\,M_7\, M_{\odot}$. In order to maintain our conclusions as general as possible, we do not specify the exact origin and location of this reconnection layer. One possible scenario is the equatorial current sheet that naturally arises in the innermost region of  magnetically-arrested disks \citep{Avara_2016,dexter_20,Ripperda:2020bpz,2021MNRAS.502.2023P,Scepi2022,2022MNRAS.513.4267N,Ripperda2022ApJ}. Another possibility is the reconnection layer adjacent to the jet boundary \citep{Nathanail_2020,Chashkina_2021,davelaar_23}, or beyond the Y-point of field lines connected to both the disk and the black hole \citep{ElMellah_2022,ElMellah_2023}. 
Regardless of its specific origin, the presence of this macroscopic reconnection layer facilitates the rapid conversion of the available magnetic energy into particle energy. A fraction of this energy can be reprocessed into hard X-rays. Traditionally, this has been associated with the Comptonization of low-energy photons from the accretion disk by a population of hot coronal electrons with $\sim 100$ keV temperature \citep{1979Natur.279..506S,1980A&A....86..121S,1991ApJ...380L..51H,1993ApJ...413..507H}. More recently, alternative scenarios have been proposed, in which X-rays are Comptonized by the bulk motions of a turbulent plasma \citep{Groselj_2023arXiv} or by the  trans-relativistic bulk motions of the stochastic chain of reconnection plasmoids \citep{Beloborodov:2017njh,Sironi:2019sxv,Sridhar:2021bvf,Sridhar:2022ojr}. We refrain from adopting a specific scenario to describe the X-ray coronal emission of NGC~1068, and use instead some general considerations applicable to accreting systems with magnetically-powered hard X-ray emission.

We consider a reconnection layer of length $L$ which we scale with the black hole gravitational radius $r_{\rm g}=G M/c^2\simeq 1.5\times 10^{12}M_7\;\rm cm$. The cross-sectional area of the ``upstream'' plasma flowing into the reconnection layer is estimated as $A_{\rm in} \sim L^2$. The characteristic thickness of the layer of reconnected plasma (the reconnection ``downstream'') is determined by the size of the largest plasmoids, which can be estimated as $w \sim \beta_{\rm rec} L$. Here, $\beta_{\rm rec}$ represents the plasma inflow speed into the reconnection layer normalized to the speed of light. Its value is also representative of the ratio between the reconnection electric field and the reconnecting magnetic field $B$.
The choice of $\beta_{\rm rec} \sim 0.1$ \citep{ComissoJPP16,CassakJPP17} is appropriate for the collisionless relativistic regime  in which the magnetization $\sigma=B^2/4\pi \rho c^2$ exceeds unity, where $\rho$ is the plasma mass density. As shown below, this is indeed our regime of interest.

In both X-ray binaries~\citep{Reig:2017art,2001MNRAS.326..417M,2000IAUS..195..153Z} and luminous AGN~\citep{Wilkins:2014caa,Petrucci:2020cda,Tripathi:2022bog}, the shape of the X-ray spectrum suggests that the optical depth for Comptonization is moderate (i.e. $\sim0.1-10$). Motivated by radiative PIC simulations~\citep{Sridhar:2021bvf,Sridhar:2022ojr}, we use the condition  $\tau_{\rm T}= n_e \sigmat w \sim 0.5$, where $\sigmat$ is the Thomson cross section and $n_e$ is the cold pair number density (henceforth, the electron density), to infer
\be
n_e\simeq 10^{12} \, \beta_{-1}^{-1}\left(\frac{LM_7}{5 r_g}\right)^{-1}\rm cm^{-3} \, ,
\label{eq:ne}
\ee
where $\beta_{-1}=\beta_\mathrm{rec}/10^{-1}$.

The strength $B$ of the reconnecting field can be obtained by assuming that the coronal hard X-ray emission comes from dissipation of magnetic energy in reconnection, as argued in \citet{Beloborodov:2017njh}. We assume that a fraction $\eta_X\sim 0.5$ of the Poynting flux crossing the reconnection plane is dissipated into Comptonized X-rays (the rest stays in electromagnetic fields, see, e.g., \citealt{sironi_15}), namely
\be \label{eq:F_X}
F_X= \eta_X \frac{\beta_{\rm rec} c}{4 \pi} B^2.
\ee
Noting that the X-ray flux is related to the intrinsic X-ray luminosity as $L_X= 2 F_X L^2$, where  the factor of two accounts for emission from both sides of the layer, the magnetic energy density is written as

\be
u_B=\frac{B^2}{8\pi}\simeq 3\times 10^7 L_{X,43}\left(\frac{L M_7}{5r_g}\right)^{-2} {\beta_{-1}^{-1}} \, \rm erg\,cm^{-3}
\ee
where $L_X=10^{43}\,L_{X,43}\, \rm erg/s$ is the bolometric intrinsic X-ray luminosity of NGC 1068. 
This implies a magnetic field
\begin{equation}
    B\simeq 3\times10^4 \left(\frac{L M_7}{5r_g}\right)^{-1} L_{X,43}^{1/2} \beta_{-1}^{-1/2}\;\mathrm{G}.
\end{equation}

Taking the pair number density from the optical depth argument presented above, we can estimate the magnetization of the system as
\be
\sigma=\frac{B^2}{4\pi n_e m_e c^2}\simeq 70\,  L_{X,43} \left(\frac{L M_7}{5r_g}\right)^{-1},
\ee
where we have assumed that ions do not appreciably contribute to the mass density, namely the proton number density is $n_p\ll n_e m_e/m_p$. As we  discuss in Section~\ref{sec:neutrino_emission}, this condition is supported by neutrino observations. As anticipated above, reconnection in a system with $\sigma\gg1$ is identified as ``relativistic''.

\section{Relativistic protons in the reconnection layer}\label{sec:non_thermal_protons}
In this section, we first discuss the physics of proton acceleration in relativistic reconnection, and the expected proton spectrum. We then describe the hierarchy of time scales (or equivalently of energy scales), accounting for proton acceleration, escape from the reconnection layer, and various cooling losses. We determine the steady-state proton spectrum and use it to predict the TeV neutrino flux.

\subsection{Spectrum of accelerated protons}
Our understanding of the physics of particle acceleration in relativistic reconnection has greatly advanced in recent years thanks to first-principles particle-in-cell (PIC) simulations \citep{werner_17,Zhang:2021akj,Zhang_2023arXiv,Chernoglazov_2023arXiv}.
Motivated by  3D PIC simulations of reconnection in the relativistic regime, we assume that, {\it in the absence of proton cooling losses}: (i) protons are accelerated linearly in time with an energization rate of approximately $dE_{p}/dt \simeq \beta_{\rm rec} e B c$, (ii) nearly all protons flowing into the reconnection layer are injected into the acceleration process, and (iii) the proton spectrum can be modeled as a broken power law with  $dn_p/dE_p\propto E_p^{-1}$ below some break energy $E_{p, \rm br}$, and $dn_p/dE_p\propto E_p^{-s}$ with $s \gtrsim 2$ above the break energy, up to a cutoff energy $E_{p,\rm max} \simeq \beta_{\rm rec} e B L$ corresponding to the maximum energy that protons can attain in the absence of cooling losses. 

The value of the post-break slope depends on various physical conditions, including the strength of the ``guide'' field (i.e., the non-reconnecting component orthogonal to the oppositely-directed fields). In fact, \citet{werner_17} used 3D PIC simulations to show that, as the guide field strength $B_g$ increases from $B_g/B=0$ to $B_g/B\sim 1$, the high-energy spectral slope increases from $s\simeq 2$ to $s\simeq 3$ (so, the spectrum gets steeper).
In the following, we deduce the post-break slope from IceCube neutrino observations, which suggest $s\sim 3$.

The characteristic energy at the spectral break is associated with the so-called proton magnetization $\sigma_p=B^2/4\pi n_p m_p c^2$, such that the break energy is of order $E_{p, \rm br} \sim \sigma_p m_p c^2$. This is equivalent to stating that the energy density in the non-thermal proton population, $u_p\sim n_p E_{p,\rm br}$, is comparable to the magnetic energy density, $u_B$. We will later show that $25$~TeV is the ballpark value for the break energy needed to explain the IceCube neutrino observations.

\subsection{Proton acceleration, escape and cooling}
\citet{Zhang:2021akj} and \citet{Zhang_2023arXiv} elucidated the physics of particle acceleration in relativistic 3D reconnection. They showed that particles gain most of their energy in the inflow (upstream) region, while meandering between the two sides of the reconnection layer.
Here, the timescale for proton acceleration $t_\mathrm{acc}\simeq E_p/\beta_\mathrm{rec} e Bc$ is of the order of
\begin{equation}
t_\mathrm{acc}\simeq 10^{-3}\frac{E_p}{25\;\mathrm{TeV}}\frac{L M_7}{5r_g}L_{X,43}^{-1/2}\beta_{-1}^{-1/2}\;\mathrm{s}.
\end{equation}

Particles leave the region of active acceleration when they get captured by one of the flux ropes/plasmoids of reconnected plasma. At that point, they are no longer actively accelerated, and they  either advect out of the system (on a timescale $t_{\rm esc}$), or they cool due to radiative losses. In the discussion that follows, we focus on the proton spectrum in the downstream region of reconnected plasma, assuming---based on PIC simulations \citep{werner_17,Zhang:2021akj,Zhang:2023lvw,Chernoglazov:2023ksi}---that the prior phase of active energization injects a broken power-law spectrum scaling as $\propto E_{p}^{-1}$ below the break and as $\propto E_{p}^{-s}$ above the break.
Protons escape/advect out of the reconnection region on a typical timescale
\begin{equation}
    t_\mathrm{esc}\simeq \frac{L}{c}\simeq 250\;\frac{L M_7}{5r_g}\;\mathrm{s}.
\end{equation}

The dominant channel of proton energy loss is photohadronic ($p\gamma$) inelastic scattering on the X-ray photon field; we discuss in App.~\ref{app:ouv} additional cooling processes which turn out to be subdominant. To compute the $p\gamma$ cooling rate we parameterize the X-ray coronal emission as follows. The X-ray energy density is given by $u_X\simeq L_X/L^2 c$. In the optically thick regime, we expect this estimate to be revised by a corrective factor proportional to $\tau_T$, but since $\tau_T\sim 1$ we neglect the exact numerical factors in what follows. Motivated by spectral modeling of NGC 1068 \citep{Bauer2015, Marinucci2016}, we adopt a power-law spectrum of photon index $-2$ between $E_{X,\mathrm{min}}=100$~eV and $E_{X,\mathrm{max}}=100$~keV, 
\begin{equation}\label{eq:X_ray_spectrum}
    \frac{{\rm d}n_X}{{\rm d}E_X}=
    \frac{L_X}{3\log(10)E_X^2 L^2 c}.
\end{equation}
The frequency range covered by the coronal spectrum is consistent with a scenario in which photons are initially injected in the optical-ultraviolet (OUV) band from the accretion disk, typically at energies even below $100$~eV, and subsequently undergo Comptonization in the corona up to $\sim 100$ keV. We emphasize though that the X-ray spectrum below about $1-2$~keV is not motivated by observations; on the other hand, as we discuss in more detail in Sec.~\ref{sec:specprot}, a different value of $E_{X,\mathrm{min}}$ would not alter our results. 

\begin{figure}
    \centering
    \includegraphics[width=0.47\textwidth]{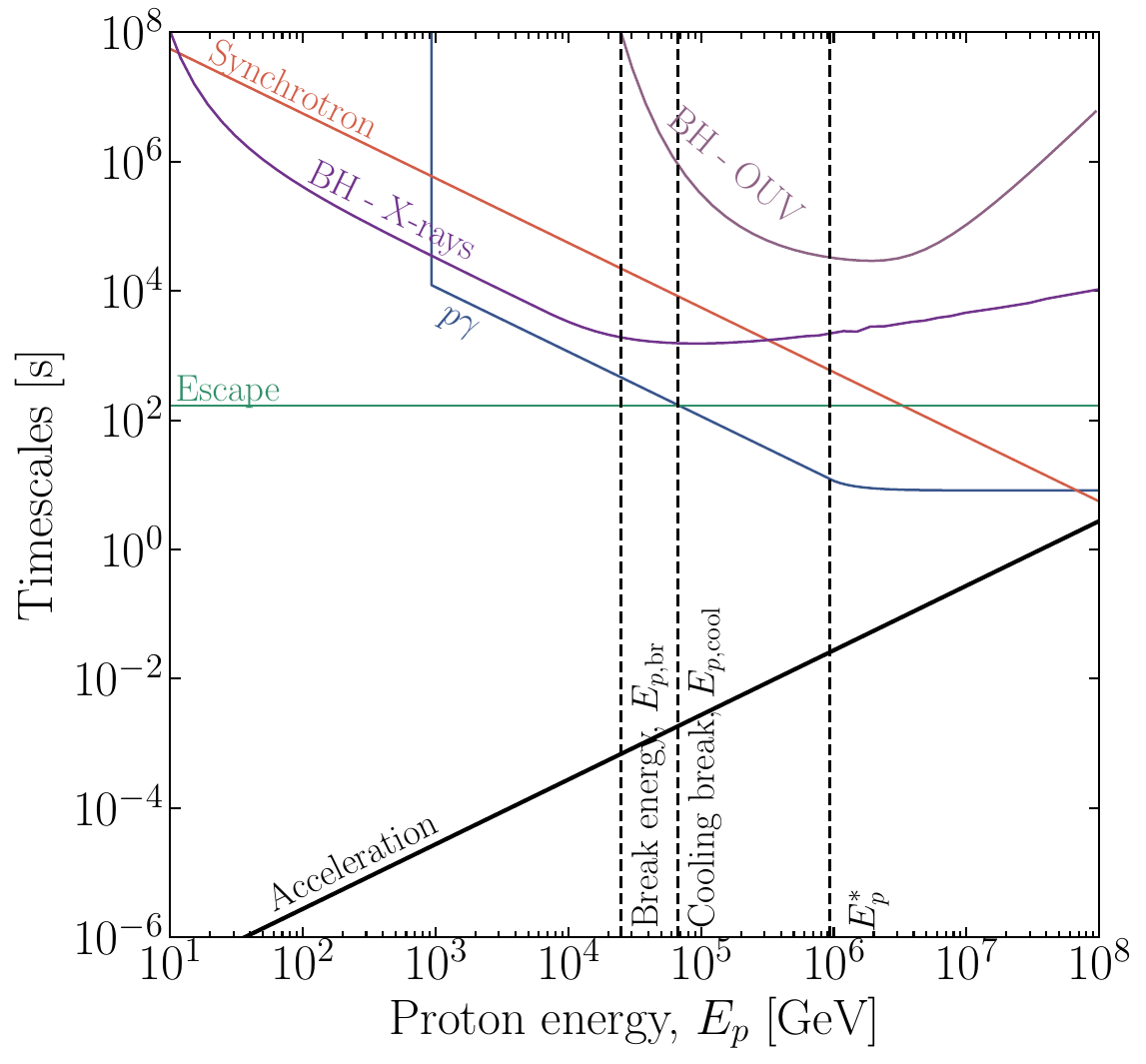}
    \caption{Timescales for proton acceleration, escape, and cooling in the reconnection region. We include synchrotron cooling, $p\gamma$ interactions with X-ray photons, and BH losses due to X-rays and optical/UV photons. We choose benchmark parameters $L=5\, r_g$, $L_{X,43}=5$,  and $M_7=0.67$. Vertical lines indicate three characteristic proton energy scales discussed in text. 
    }
    \label{fig:timescales}
\end{figure}

 We assume an effective cross section that is constant in energy, $\hat{\sigma}=\sigma_{p\gamma}K_{p\gamma}=70$~$\mu$b, for interaction energies above  the center-of-mass energy threshold $\epsilon_r=390$ (in units of $m_ec^2$), where $\sigma_{p\gamma}$ is the $p\gamma$ total cross section, and $K_{p\gamma}$ is the proton inelasticity \citep{Atoyan:2002gu, Dermer:2003zv}. Using the X-ray spectrum in Eq.~\ref{eq:X_ray_spectrum}, we estimate the $p\gamma$ energy-loss timescale as 
\begin{eqnarray}
    t_{p\gamma}&=&\frac{9\log(10)L^2\ \epsilon_r \, m_e c^2}{4\hat{\sigma} L_X (E_p/m_pc^2)}\\ \nonumber &=&
     4.8\times 10^3\;\left(\frac{L M_7}{5r_g}\right)^2 \frac{25\;\mathrm{TeV}}{E_p} L_{X,43}^{-1}\;\mathrm{s},
\end{eqnarray}
for proton energies $E_p<E_p^*$, which is defined as
\begin{equation}
    E^*_p=\frac{\epsilon_r \, m_p m_e c^4}{2E_{X,\mathrm{min}}}\simeq 930\;\mathrm{TeV}.
\end{equation}
In contrast, for $E_p>E^*_p$ the cooling timescale flattens to the asymptotic energy-independent value 
\begin{eqnarray}
   t_{p\gamma}&=&\frac{3\log(10)L^2 \epsilon_r \, E_{X,\mathrm{min}}}{\hat{\sigma} L_X}\\ \nonumber &=&
     87\;\left(\frac{L M_7}{5r_g}\right)^2 L_{X,43}^{-1}\;\mathrm{s},
\end{eqnarray}
since the scattering does not happen anymore at threshold for pion production, and becomes dominated by multi-pion interactions with the whole X-ray spectrum. 

Another relevant energy scale is obtained by equating the cooling timescale with the escape/advection timescale
\begin{equation}
    E_{p,\mathrm{cool}}=490\;\mathrm{TeV}\;L_{X,43}^{-1}\frac{L M_7}{5 r_g}.
    \label{eq:Epcool}
\end{equation}
Above $E_{p,\mathrm{cool}}$, protons cool much faster than they can escape the reconnection layer, and therefore lose most of their energy while being inside of it.

The maximum energy to which protons can be accelerated is set by the equality between the acceleration and the dominant energy-loss timescale. At very high energies, this can either be synchrotron (estimated in App.~\ref{app:ouv}) or $p\gamma$, yielding (first and second term in the square bracket, respectively)
\begin{eqnarray}
    E_{p,\mathrm{rad}}\simeq 380\;\left(\frac{L M_7}{5 r_g} \right)^{1/2}L_{X,43}^{-1/2} \beta_{-1}^{3/4}\\ \nonumber\mathrm{min}\left[1,5.6\left(\frac{L M_7}{5 r_g} \right)^{1/2}L_{X,43}^{-1/2}\beta_{-1}^{-1/4}\right]\;\mathrm{PeV}.
\end{eqnarray}

Fig.~\ref{fig:timescales} collects the energy-dependent timescales we have introduced, for a benchmark choice of parameters representative of the likely conditions in NGC 1068 (see Sec.~\ref{sec:ngc1068}). We also show the synchrotron energy-loss timescale, and the BH energy-loss timescales due to X-ray and OUV photons separately, as discussed in App.~\ref{app:ouv}. We can grasp directly the main features discussed in the text: the acceleration timescale is much faster than all the other timescales up to hundreds of PeV. However, protons cool faster than the escape timescale, due to $p\gamma$ interactions, already very close to the break energy.

\subsection{Steady-state proton spectrum}\label{sec:specprot}
Here we consider the proton spectrum in the downstream region of reconnected plasma, assuming that the prior phase of active acceleration in the upstream injects a broken power law scaling as $\propto E_{p}^{-1}$ below the break and as $\propto E_{p}^{-s}$ above the break.
When accounting for cooling losses, the  steady-state proton spectrum will be composed of several power-law segments that depend on the hierarchy of the proton energy scales. For $E_{p,\mathrm{br}} <  E_{p,\mathrm{cool}} < E^*_p < E_{p,\mathrm{rad}}$, as shown in Fig.~\ref{fig:timescales}, we expect 

\begin{equation}
    \frac{dn_p}{dE_p}\propto
    \begin{cases}        
    \left(\frac{E_p}{E_{p,\mathrm{br}}}\right)^{-1},  E_p<E_{p,\mathrm{br}}\\
    \left(\frac{E_p}{E_{p,\mathrm{br}}}\right)^{-s},  E_{p,\mathrm{br}}<E_p<E_{p,\mathrm{cool}}\\
    \left(\frac{E_{p,\mathrm{cool}}}{E_{p,\mathrm{br}}}\right)^{-s}\left(\frac{E_p}{E_{p,\mathrm{cool}}}\right)^{-s-1},  E_{p,\mathrm{cool}}<E_p<E^*_{p}\\
    \frac{E_{p,\mathrm{br}}^s E_{p,\mathrm{cool}}}{(E_p^{*})^{s+1}}\left(\frac{E_{p}}{E^*_{p}}\right)^{-s},   E^*_{p}<E_p<E_{p,\mathrm{rad}}.
    \end{cases}    
\label{eq:npdiff}
\end{equation}

Notice that the break at $E^*_p$, which is determined by the lower energy of the X-ray target $E_{X,\mathrm{min}}$, is typically much above the peak of the proton spectrum, and therefore does not impact the neutrino production close to the peak. Therefore, a lower value for $E_{X,\mathrm{min}}$ would not significantly affect our results.

The overall normalization of the proton spectrum is determined by the condition of near-equipartition between magnetic field energy density and proton energy density. Specifically, we normalize the spectrum by assuming that the proton energy density is $u_p=\eta_p u_B$. In the absence of proton cooling losses, PIC simulations indicate that $\eta_p=(u_p/u_B)_{\rm uncool}\gtrsim 0.1$ \citep{french_23,totorica_23,Com23}. In what follows, we consider the special case where protons that carry most of the energy (i.e., those at $E_{p, \rm br}$) are marginally fast-cooling, namely $E_{p,\mathrm{cool}}\gtrsim E_{p,\mathrm{br}}$, and use $\eta_p= 0.1\eta_{p,-1}$ as a typical value. In App.~\ref{app:strong_cooling}, we generalize our results to the fast-cooling regime where $E_{p,\mathrm{cool}} \ll E_{p,\mathrm{br}}$. 
 
If we take the energy hierarchy $E_{p,\mathrm{br}} \sim   E_{p,\mathrm{cool}}\ll E^*_p \ll E_{p,\mathrm{rad}}$, with post-break slope of $s=3$ (motivated by the neutrino observations), we can compute the total proton number density,
\begin{eqnarray}\label{eq:proton_number_density}
    n_p&\simeq& \frac{2\eta_p u_B}{3E_{p, \mathrm{br}}}\left[\frac{1}{3}+\log\left(\frac{E_{p,\mathrm{br}}}{E_{p,\mathrm{min}}}\right)\right] \approx  \frac{20\eta_p u_B}{3 E_{p, \mathrm{br}}} \nonumber  \\ 
      &\simeq& 5\times 10^5 \frac{L_{X,43}\eta_{p,-1}}{\beta_{-1}}\left(\frac{L M_7}{5r_g}\right)^{-2}\frac{25\;\mathrm{TeV}}{E_{p,\mathrm{br}}}\;\mathrm{cm}^{-3}  
\end{eqnarray}
where we have replaced the square bracket by its typical order of magnitude value $10$ for $E_{p,\mathrm{br}}\sim 25$~TeV (i.e. typical value needed to reproduce the IceCube neutrino measurements).  Eq.~(\ref{eq:proton_number_density}) confirms our initial assumption that 
$n_p m_p \ll n_e m_e$, see Eq.~(\ref{eq:ne}). Charge neutrality therefore requires the largest part of the lepton population in the coronal region to be made of electron-positron pairs.

\subsection{Neutrino emission}\label{sec:neutrino_emission}
Photohadronic collisions inject charged pions, which subsequently decay to neutrinos. We estimate the production rate of neutrinos and anti-neutrinos of all flavors as
\begin{equation}
\label{eq:16}
    E_\nu^2 Q_\nu =L^3 \beta_\mathrm{rec} \, \left[K_\nu\frac{E_p^2 dn_p/dE_p}{t_{p\gamma}}\right]_{E_p=20 E_\nu},
\end{equation}
where $Q_{\nu}$ is {the rate of neutrino production} differential in neutrino energy, $\beta_\mathrm{rec}L^3$ is the volume of the reconnection region, and the factor $K_\nu$ is the fraction of proton energy converted to neutrinos. For interactions at  $\Delta$-resonance $K_\nu=1/4$, while for multi-pion interactions $K_\nu=1/2$. For the target X-ray  spectrum we are considering, using the two-bin parameterization for the $p\gamma$ interaction rate of~\cite{Dermer:2003zv}, we find that around $E_p\simeq E_{p,\mathrm{br}}$ a reasonable approximation is $K_\nu\simeq 0.35$.
We assume here that each neutrino carries on average an energy $E_\nu=E_p/20$. 

Since $t_{p\gamma}\propto E_p^{-1}$ for $E_p<E_p^*$, the neutrino spectrum is hardened by one power of energy as compared to the proton spectrum, whereas it has the same slope as the proton spectrum for $E_p>E_p^*$. Taking the post-break proton slope to be $s=3$ as in Eq.~\ref{eq:npdiff}, we expect a broken power law with $E_\nu^2 Q_\nu \propto E_\nu^2$ for $E_\nu<0.05 E_{p,\mathrm{br}}$; $E_\nu^2 Q_\nu\propto E_\nu^0$ for $0.05 E_{p,\mathrm{br}}<E_\nu<0.05 E_{p,\mathrm{cool}}$; and $E_\nu^2 Q_\nu \propto E_\nu^{-1}$ for $E_\nu>0.05 E_{p,\mathrm{cool}}$ (since the proton spectrum is steepened by photohadronic losses to $dn_p/dE_p\propto E_p^{-4}$), see also Eq.~\ref{eq:npdiff}. 
Thus, for our benchmark value $s=3$ of the post-break proton slope, the neutrino spectrum is consistent with IceCube observations. 
The peak value of $E_\nu^2 Q_\nu$, which lies in the energy range $0.05 E_{p,\mathrm{br}} - 0.05 E_{p,\mathrm{cool}}$, is
\begin{eqnarray}
    \label{eq:peak-nu}
    &&(E_\nu^2 Q_\nu)^\mathrm{pk}=\frac{2K_\nu\eta_p E_{p,\mathrm{br}}L_X^2 \hat{\sigma}}{27\log(10)\eta_X \epsilon_r m_e m_p c^5 L}\simeq\\ \nonumber && 5.9\times 10^{39}\eta_{p,-1} L_{X,43}^2 \frac{5 r_g}{L M_7}  \frac{E_{p,\mathrm{br}}}{25\;\mathrm{TeV}}\;\mathrm{erg/s}
\end{eqnarray}
For $L_{X,43}\gtrsim5-10$, this is comparable with the required luminosity needed to explain the IceCube flux, provided that $E_{p, \rm br}$ is in the range $50-100$~TeV.

\subsection{Proton-mediated pair enrichment}
So far, we have left unspecified the origin of the leptonic population, whose number density we can infer from the Compton opacity. A thrilling possibility is that the proton population itself is responsible for maintaining a steady 
density of cold pairs in the coronal region. MeV photons, which are primarily emitted by Bethe-Heitler relativistic pairs (as we will show in Sec.~\ref{sec:ngc1068}), as well as cascaded from $p\gamma$-produced gamma rays, inject pairs almost at rest after their attenuation. Because of the hadronic origin of MeV photons, we argue that the typical luminosity in the MeV band is comparable with the neutrino luminosity (see also Sec.~\ref{sec:ngc1068}). An  order-of-magnitude estimate of the MeV photon energy density is
\begin{equation}
    u_\mathrm{MeV}\sim \frac{(E_\nu^2 Q_\nu)^\mathrm{pk}}{L_X}u_X\simeq \frac{(E_\nu^2 Q_\nu)^\mathrm{pk}}{L^2 c}
\end{equation}
and their number density is $n_\mathrm{MeV}\sim u_\mathrm{MeV}/1\;\mathrm{MeV}$. While this is only an order-of-magnitude estimate, it captures the expected scaling with the energy injected by  hadrons; we validate it by comparing with our numerical simulations in Sec.~\ref{sec:ngc1068}. Finally, we can estimate the amount of pairs produced by MeV photon-photon attenuation as
\begin{eqnarray}\label{eq:produced_pairs}
    &&n_e^{\gamma\gamma}\sim 0.1\sigma_T n_\mathrm{MeV}^2 L\\ \nonumber &&\simeq  2.5\times10^6\;L_{X,43}^4 \eta_{p,-1}^2\left(\frac{L M_7}{5 r_g}\right)^{-5}\left(\frac{E_{p,\mathrm{br}}}{25\;\mathrm{TeV}}\right)^2\;\mathrm{cm}^{-3}.
\end{eqnarray}
For $L_{X,43}\gtrsim 5-10$, the attenuation of $\sim$MeV photons can therefore lead to a pair population that constitutes a sizeable fraction of the pair density inferred from the Compton opacity in Eq.~\ref{eq:ne}. The above estimate should be taken as a lower limit, because we have only considered pair production via attenuation of $\sim$MeV photons. We have not included, e.g., the contribution of relativistic pairs---produced in photohadronic interactions and from the attenuation of GeV-TeV gamma-ray photons---that cool down to Lorentz factors $\gamma_e\sim 1$.

\subsection{Neutrino emission from NGC 1068}\label{sec:ngc1068}

\begin{figure}
    \centering
    \includegraphics[width=0.47\textwidth]{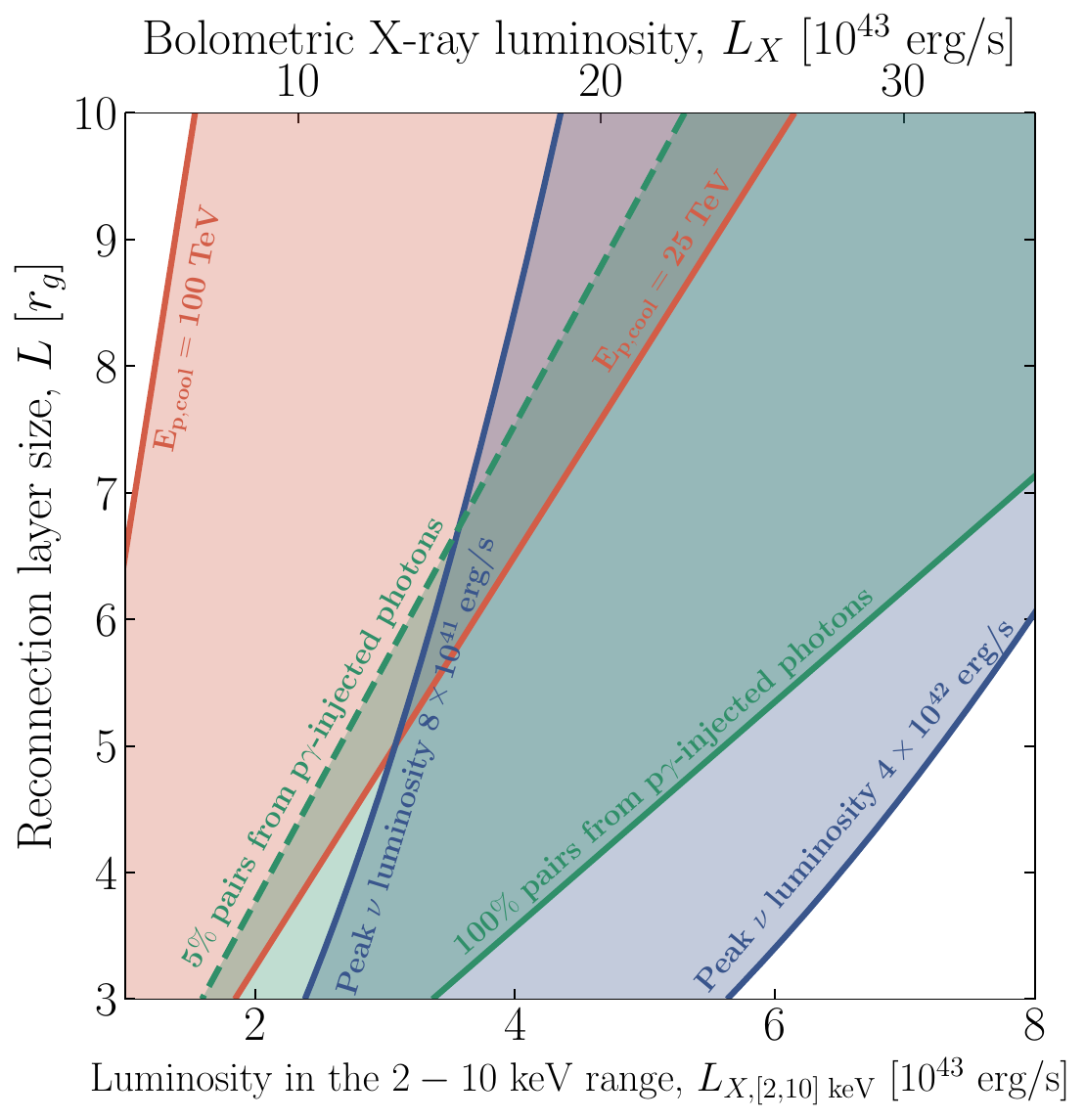}
    \caption{Regions of the $L-L_{X, [2,10]~\rm keV}$ parameter space leading to a neutrino signal from NGC~1068 consistent with IceCube observations. The axis at the top shows the bolometric intrinsic X-ray luminosity $L_X$, which we use in our equations. We show in red the requirements on the cooling break (Eq.~\ref{eq:Epcool}) and in blue the constraints on the peak neutrino luminosity (Eq.~\ref{eq:peak-nu}). We show in green the region where the pairs injected by the attenuation of hadronically-produced MeV photons
    have a number density from $5\%$ to $100\%$ of the pair density inferred from the Compton opacity. We use benchmark values $M_7=0.67$, $\eta_p=0.05$, $E_{p,\mathrm{br}}=25$~TeV.}
    \label{fig:example_regions}
\end{figure}
We propose that reconnection-powered coronal neutrino emission is the dominant source of the neutrino flux observed by IceCube in coincidence with NGC 1068. Results are shown for a luminosity distance $d_L = 10.1$~Mpc \citep[see][and Padovani et al., in prep.]{Tully_2008} and a black hole mass $M_7=0.67$ \citep[see][and Padovani et al., in prep.]{Greenhill1996}. 

The first requirement is that the neutrino luminosity matches the observed all-flavor luminosity observed by IceCube. 
Assuming a peak neutrino energy of $1.25$~TeV, corresponding to a proton break energy $E_{p,\mathrm{br}}\simeq 25$~TeV, the all-flavor neutrino luminosity is  $(0.8-4)\times10^{42}$~erg/s (based on the best-fit value and $95\%$ confidence region shown in Fig.~4 of \cite{IceCube-NGC1068}, after accounting for the difference in the luminosity distance used). The second requirement is that the neutrino spectrum observed by IceCube must be sufficiently soft, which can be accommodated in our model if the post-break proton slope is $s \sim 3$ and simultaneously the cooling break $E_{p,\mathrm{cool}}$ is not much larger than the break energy $E_{p,\mathrm{br}}\simeq 25\,\rm{TeV}$.  In this way, the neutrino spectrum is not hardened significantly by the energy dependence of the $p\gamma$ efficiency ($t_{p\gamma}\propto E_p^{-1}$ in Eq.~\ref{eq:16}), and the reconnection layer acts almost as a proton calorimeter. Therefore, our second constraint is that  $E_{p, \rm cool}\gtrsim E_{p, \rm br}\simeq25$~TeV. We take $E_{p, \rm cool}$ to lie between 25~TeV and 100 TeV (the fast-cooling case $E_{p,\mathrm{cool}}\ll E_{p,\mathrm{br}}$ is discussed in more detail in App.~\ref{app:strong_cooling}.) 

\begin{figure*}
\centering
    \includegraphics[width=0.7\textwidth]{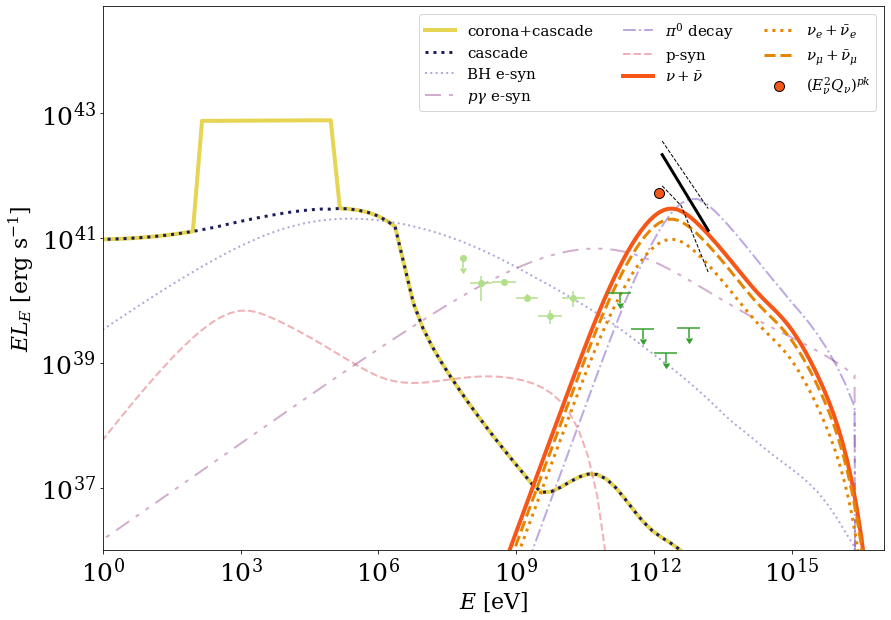}
    \caption{Spectral energy distribution of photons (thick solid yellow line) and all-flavor neutrinos (thick solid orange line) from the reconnection region. Single-flavor neutrino spectra, not including mixing (i.e., flavor oscillations), are also shown (dashed and dotted orange lines). For comparison, we show the estimated peak neutrino luminosity in Eq.~\ref{eq:Lnu_pk_sphere} (orange circle). The non-thermal cascade spectrum is overplotted with a dotted black line. Other lines indicate emission from different processes (see inset legend) before accounting for $\gamma \gamma$ pair production. Gamma-ray data by Fermi-LAT \citep{Fermi_LAT_obs} and upper limits by MAGIC \citep{MAGIC-UL-NGC1068} are also included (light and dark green markers respectively). Solid and dashed black lines show the best-fit all-flavor IceCube neutrino spectrum with the 68\% confidence interval, adopted from \citet{IceCube-NGC1068}.} \label{fig:sed}
\end{figure*}

Fig.~\ref{fig:example_regions} shows the regions of parameter space where both requirements are simultaneously satisfied, according to our analytical estimates. We consider as independent parameters the size $L$ of the reconnection layer and the X-ray luminosity in the $2-10$~keV range; while the latter is in principle observed, there is a relatively large uncertainty on its value $L_{X,[2-10]\;\mathrm{keV}}=3^{+3}_{-2}\times 10^{43}$~erg/s \citep[see][and Padovani et al., in prep; we rescaled their luminosity to account for the different choice of luminosity distance]{Marinucci2016}, motivating the range chosen for the figure. The axis at the top shows the bolometric (0.1~keV -- 100 keV) intrinsic X-ray luminosity $L_X$, which we have used in our analytical estimates. We also show in green the region where the pairs produced  by the attenuation of MeV photons according to Eq.~\ref{eq:produced_pairs} can contribute a sizable fraction, if not the entirety, of the pair population inferred from the Compton opacity argument. The allowed parameter space provides a coherent explanation for both the neutrino signal and the moderate optical depth of coronal regions.

To confirm our analytical estimates, we have also performed numerical calculations of the neutrino and gamma-ray production within a spherical coronal region of size $R$; we describe in Appendices~\ref{app:spherical_geometry} and~\ref{app:numerical_approach} the relation between the spherical geometry of our numerical calculations and the planar one used in our analytical estimates, as well as the main properties of the numerical code we used. We show the resulting neutrino spectral energy distribution in Fig.~\ref{fig:sed} (orange solid line), whose peak is in reasonable agreement with the analytical expectation (orange circle)\footnote{A comparison of the numerically-computed steady-state proton distribution with the analytical expression of Eq.~\ref{eq:npdiff} can be found in App.~\ref{app:numerical_approach}.}. Our calculations do not account for neutron-photon interactions; we have separately evaluated them numerically, finding that they can account for a $50\%$ increase in the neutrino flux shown in Fig.~\ref{fig:sed}. 

The large opacity to photon-photon pair production of the coronal region suppresses the gamma-ray spectrum at energies $\gtrsim 1$~MeV, making it consistent with the observations.
Synchrotron radiation of Bethe-Heitler pairs is an important source of $\sim$~MeV photons, which produce pairs with $\gamma_e \sim 1$. Attenuation of more energetic photons arising from $p\gamma$ interactions leads to the production of relativistic secondary electrons and positrons, which cool down to $\gamma_e \sim 1$ within a dynamical time due to the strong synchrotron losses. As a result, attenuation of gamma-ray photons produced from photohadronic interactions provides another channel for injecting  cold pairs in the system. With our numerical code, we find that the density of pairs resulting from the attenuation of hadronically-produced photons is $n_{e, \rm cold} \sim 3 \times 10^{11}$~cm$^{-3}$, which leads to $\tau_T = \sigma_T\, R\,  n_{e, \rm cold}\sim 0.3$. {This numerical value is also consistent with the analytical estimate computed for a spherical geometry, see Eq.~\ref{eq:negg-sphere}}. 

{We note that the radiative calculations we presented have some limitations.} First, cold pairs are ``passive'' (i.e., non-emitting) particles in our code, since radiative processes of non-relativistic electrons (e.g., cyclotron radiation and Compton scattering) are not implemented. Second, we do not account for the energization of leptons due to magnetic reconnection or their synchrotron emission, which might be important for the low-energy end of the photon spectrum. For this reason, we only show the photon spectrum down to 1~eV, and consider the X-ray coronal emission given (i.e., no interaction between cold pairs and X-ray photons is considered).

\section{Summary and Discussion}\label{sec:discussion}
We propose that protons energized by magnetic reconnection in a highly magnetized, pair-dominated plasma interact with hard X-rays from the corona and produce the high-energy neutrinos observed by IceCube from the Seyfert galaxy NGC~1068. The key ingredient of this study---grounded on recent 3D PIC simulations of relativistic reconnection---is that the X-ray coronal emission, the energy density of reconnecting magnetic fields, and the relativistic proton population are all connected with each other. 

If hard X-ray photons are produced via Comptonization of lower energy photons by pairs in the reconnection region, as previously proposed, then the energy density of up-scattered photons will be a fraction (of order $\sim 0.1$) of the  energy density of reconnecting magnetic fields. Protons accelerated in the reconnection region will also carry a fraction $\sim0.1$ of the magnetic energy density. To avoid an ``energy crisis'', the hard $\propto E_p^{-1}$ proton spectrum established at low energies must soften ($\propto E_p^{-s}$, with $s>2$) above a characteristic break energy $E_{p, \rm br}$. From the peak of the observed neutrino flux, we infer $E_{p, \rm br}\simeq 25~$TeV. 
Under these conditions, the neutrino emission
can be fully determined by only two parameters: the size of the reconnection region, and the X-ray luminosity of the corona (which is observationally constrained, but with large error bars).

In our scenario, the proton break energy happens to be just below the energy at which photohadronic cooling becomes faster than the proton escape/advection from the reconnection layer. This implies that most of the energy in the accelerated protons is lost to neutrinos and gamma rays. While the latter are not able to escape the optically-thick reconnection layer, and are cascaded down to lower energies, neutrinos carry away a significant fraction of the proton energy, explaining the relatively large flux observed by IceCube. Our scenario naturally leads to a robust scaling of the peak neutrino luminosity with black hole mass and X-ray luminosity, i.e., $(E^2_\nu Q_\nu)^{\rm pk}\propto M^{-1} L_X^2$, which comes from the direct connection between X-ray, magnetic, and non-thermal proton energy density. In the case of fast-cooling, with $E_{p,\mathrm{cool}}\ll E_{p,\mathrm{br}}$, the scaling with $L_X$ becomes linear with no dependence on black hole mass and layer size, since the reconnection layer becomes calorimetric. 

Intriguingly, we also find that the pair population induced by the cascade initiated by $p\gamma$ interactions and by the attenuation of Bethe-Heitler-produced MeV photons is comparable with the one inferred from the Compton optical thickness of the corona. Even though this result is not a strict requirement, i.e., the bulk of the pairs might have a different origin, it is a natural outcome of our model, at least for the parameters of NGC~1068. We are therefore prompted to speculate that the non-thermal proton population may play an important role also in maintaining the large lepton density in AGN coronae. A more thorough dynamic study of the interplay between protons and X-ray coronal photons is yet to be carried out.

Based on our findings, reconnection layers in AGN coronae are magnetically-dominated (i.e., the magnetic energy density is much larger than the plasma rest-mass energy density) and significantly pair-enriched, to the point that the upstream mass density is mostly contributed by electron-positron pairs. Equivalently, if the magnetization $\sigma$ is normalized to the pair mass density, we have $1\ll\sigma\ll\sigma_p$, where the ``proton magnetization'' is (see Eq.~\ref{eq:proton_number_density}) 
\begin{equation}
    \sigma_p=\frac{B^2}{4\pi n_p m_p c^2}\sim \frac{3 E_{p,\mathrm{br}}}{10 \eta_p m_p c^2}\simeq 8\times 10^4\;\eta_{p,-1}^{-1}\frac{E_{p,\mathrm{br}}}{25\;\mathrm{TeV}}.
\end{equation}
The low value of the proton number density (equivalently, the high $\sigma_p$) seems to be somewhat needed for the observability of the neutrino signal. A much higher proton density would correspond to a much lower break energy (at fixed magnetic energy density), so the neutrino signal would peak in an energy range unobservable by IceCube.

Interestingly, the three scenarios  presented in Section \ref{sec:global_properties} for the formation of a large-scale coronal reconnection layer are likely consistent with a pair-dominated, proton-poor composition (in GRMHD simulations, the particle density in the layer is found to be dominated by artificially-injected density floors, rather than by the physical density of the initialized electron-proton torus)\footnote{This can also be the case for reconnection layers at the jet/wind boundary \citep{Nathanail_2020,Chashkina_2021,davelaar_23}, if reconnection occurs because of jet field lines winding up onto themselves due to the underlying velocity shear, see \citet{sironi_21}.}. To the best of our knowledge, 3D PIC simulations in this regime ($1\ll\sigma\ll\sigma_p$) have yet to properly quantify how the post-break proton slope $s$ depends on the guide field strength. \citealt{werner_17} demonstrated that the high-energy spectral slope is steeper for stronger guide fields---with $s\simeq 3$ for a guide field comparable to the reconnecting field---but their simulations adopted  an electron-positron composition. If our observationally-motivated choice of $s=3$ is indeed to be attributed to a guide field comparable to the reconnecting field, then the reconnection layer at the jet boundary is the most likely scenario \citep[e.g.,][]{sironi_21}. But in general, whether a relatively strong guide field is attained in the three scenarios presented in Section~\ref{sec:global_properties} is yet to be robustly assessed.  As far as current simulations are concerned~\citep{Ripperda2022ApJ}, equatorial current sheets do not seem to host a guide field, although this statement has been verified only for a relatively large black hole spin. Finally, for the Y-point of field lines connected to the disk and the black hole, \cite{ElMellah_2022} exhibits a low guide field, but the field geometry here is constrained by the boundary conditions to be anchored to the accretion disk. Future simulations, where the disk is not merely a boundary condition but is self-consistently evolved, will be needed to study the properties of the reconnection layer in this case. Such studies will be essential to support our observationally-motivated choice of $s\simeq 3$ for NGC~1068.

A direct extrapolation of our results for NGC~1068 to other AGN might suggest that neutrino emission from coronal regions requires the presence of a jet. However, the requirement of a significant guide field, which could be realized in a jet/disk boundary layer, only holds for NGC~1068, for which a  soft neutrino spectrum is observed. Harder neutrino spectra can be produced in equatorial current sheets at AGN without jets. In conclusion, steep neutrino spectra ($s>2$) from coronal regions of AGN would typically require reconnection layers in the jet/disk boundary where stronger guide fields may be found, otherwise, a jet is not needed. Future discoveries of other neutrino-emitting Seyfert galaxies with IceCube and KM3Net will critically test our proposed scenario. 
 
\textit{Note added. -} While we were finalizing this work, the paper by \citet{2023arXiv231015222M} appeared. The two works are
independent and complementary. Some arguments are similar---most notably, by assuming that Comptonization happens in a magnetically-dominated corona, one can infer the coronal density and field strength directly from X-ray observations---but there are also important differences. \citet{2023arXiv231015222M} invoked acceleration in reconnection as the injection stage for further acceleration by turbulence. Instead, we propose a simpler, better constrained model based only on reconnection. We also supplement our analytical estimates with numerical calculations of the full electromagnetic cascade, highlighting the importance of reconnection-accelerated protons  for  sustaining the population of electron-positron pairs that makes the corona moderately Compton thick.

\section*{Acknowledgements} 
This work was supported by a grant from the Simons Foundation (00001470, to L.S.). We are grateful to Paolo Padovani for useful comments on the draft.
L.S. acknowledges support from DoE Early Career Award DE-SC0023015. This research was facilitated by the Multimessenger Plasma Physics Center (MPPC), NSF grant PHY-2206609 to L.S. 
M.P. acknowledges support from the MERAC Fondation through the project THRILL and from the Hellenic Foundation for Research and Innovation (H.F.R.I.) under the ``2nd call for H.F.R.I. Research Projects to support Faculty members and Researchers'' through the project UNTRAPHOB. DFGF is supported by the Villum Fonden under Project No.\ 29388 and the European Union's Horizon 2020 Research and Innovation Program under the Marie Sk{\l}odowska-Curie Grant Agreement No.\ 847523 ``INTERACTIONS.''
L.C acknowledges support from the NSF grant PHY-2308944 and the NASA ATP grant 80NSSC22K0667.
EP acknowledges support from the Villum Fonden (No.~18994) and from the European Union’s Horizon 2020 research and innovation program under the Marie Sklodowska-Curie grant agreement No. 847523 ‘INTERACTIONS’.
EP was also supported by Agence Nationale de la Recherche (grant ANR-21-CE31-0028).

\bibliographystyle{aasjournal}
\bibliography{References}

\appendix

\section{Bethe-Heitler energy losses}\label{app:ouv}

In the main text, we have considered as a dominant loss channel the photohadronic interaction of reconnection-accelerated protons onto the X-ray photon field. Here we consider additional channels of energy loss. 

Other works on the coronal emission, e.g.~\cite{Murase:2019vdl}, have pointed at Bethe-Heitler (BH) interactions with the optical-ultraviolet (OUV) photon field as the dominant energy loss channel. The reason for our different conclusion is the location of the neutrino production site, which in our scenario, is a reconnection region formed much closer to the central black hole. While the X-ray field is directly produced within the reconnection region, the OUV field 
is thought to be produced in a thin accretion disk lying at larger radii, $\sim 10-50\,r_g$, and therefore the energy density of optical photons is geometrically diluted within the region of interest. In this section, we explicitly verify these statements.

Let us first consider BH scattering on the X-ray target, to show that it is generally subdominant as compared to $p\gamma$ interactions. The BH energy loss timescale of a proton with Lorentz factor $\gamma_p$ due to interactions with an isotropic photon field of differential number density distribution
$dn_\gamma(E_\gamma)/dE_\gamma$ reads 
\begin{equation}
    t_{\mathrm{BH}}^{-1}(\gamma_p)=\sigma_T\left(\frac{m_e}{m_p}\right)\frac{m_e c^3}{\gamma_p^2}\int_{\gamma_p^{-1}}^{+\infty}\frac{d\epsilon}{\epsilon^2} \frac{dn_\gamma(\epsilon m_e c^2)}{dE_\gamma}\Phi(2\gamma_p \epsilon),
\end{equation}
where
\begin{equation}
    \Phi(x)=\int_2^{x} d\overline{\epsilon}\; \overline{\epsilon}\frac{\sigma_{\phi e}(\overline{\epsilon})}{\sigma_T\sqrt{1+2\overline{\epsilon}}}
\end{equation}
and we take the BH cross section $\sigma_{\phi e}(\overline{\epsilon})$ from the numerical fits of~\cite{Stepney_rates}. For the X-ray photon target density given by Eq.~\ref{eq:X_ray_spectrum}, which we report here for convenience
\begin{equation}
    \frac{dn_X}{dE_X}=\frac{L_X}{3\log(10)E_X^2 L^2 c}, \, E_{X,\mathrm{min}}<E_X<E_{X,\mathrm{max}},
\end{equation}
we obtain the energy loss rate as
\begin{equation}
    t_{\mathrm{BH},X}^{-1}(\gamma_p)=\frac{\sigma_T L_X}{3\log(10)m_pL^2c^2}\frac{\xi_X(\gamma_p)}{\gamma_p^2}, 
\end{equation}
where $\xi_X(\gamma_p)$ is the dimensionless function defined by
\begin{equation}
    \xi_X(\gamma_p)=\int_{\mathrm{min}\left(\gamma_p^{-1},\frac{E_{X,\mathrm{min}}}{m_e c^2}\right)}^{\frac{E_{X,\mathrm{max}}}{m_e c^2}}\frac{d\epsilon}{\epsilon^4}\Phi(2\gamma_p\epsilon).
\end{equation}
For comparison, the photohadronic timescale for $\gamma_p<\gamma_p^*=E_p^*/m_pc^2$ is
\begin{equation}
t_{p\gamma}^{-1}(\gamma_p)=\frac{4\hat{\sigma}L_X}{9\log(10) L^2 \epsilon_r m_e c^2}\gamma_p.
\end{equation}
The ratio between these two timescales does not depend on the X-ray luminosity or the size of the region; noting that the function $\xi_X(\gamma_p)/\gamma_p^3$ has a maximum value of about $2\times 10^{-4}$, we easily verify that the ratio between the two timescales is bounded by
\begin{equation}
    \frac{t_{p\gamma}}{t_{\mathrm{BH},X}}\lesssim 0.3.
\end{equation}
Therefore, BH energy losses from the X-ray field are at most comparable and slightly subdominant with respect to the photohadronic interactions off the same target field for protons with $E_p \le E_p^*$. Nevertheless, the energy lost via $p\gamma$ interactions is divided into 3 particle species (neutrinos, gamma-rays, and pairs), while the proton energy lost via Bethe-Heitler is channeled to pairs alone. Therefore, the emission of Bethe-Heitler pairs and neutrinos may be comparable, as shown in Fig.~\ref{fig:sed} \citep[see also,][]{PM2015}. 

The additional target field in the region of interest is the OUV field, which 
is produced mostly as 
a multi-temperature spectrum from the geometrically thin, optically thick accretion disk at larger radii. First of all, we provide an estimate of how far away from the central black hole the OUV is mostly produced. We consider an OUV-integrated luminosity $L_\mathrm{OUV}=\phi L_X$, where the dimensionless factor is typically $\phi\simeq 10$  \citep[][and references therein]{Murase_2020}. 
Assuming an accretion efficiency $\eta_a\simeq 0.1$, this points to a mass accretion rate
\begin{equation}
\label{eq:mdot}
    \dot{M}\simeq\frac{\phi L_X}{\eta_{\rm acc} c^2}.
\end{equation}
The effective temperature of the produced OUV field as a function of radius is determined by the standard Shakura-Sunyaev theory of geometrically thin accretion disks. The typical energy of the OUV field relevant for BH interactions is of order of $10$~eV, so we consider an effective temperature $T_\mathrm{OUV}=3.33$~eV; the corresponding radius is
\begin{equation}
    r_\mathrm{OUV}=\left(\frac{3GM \dot{M}}{8\pi\sigma_\mathrm{SB} T_\mathrm{OUV}^4}\right)^{1/3}\simeq 75 r_g L_{X,43}^{1/3} M_7^{-2/3}
\end{equation}
where $\sigma_\mathrm{SB} $ is the Stefan-Boltzmann constant. This distance is typically much larger than the one considered for the reconnection region in our scenario, which is already suggestive of the subdominant nature of the OUV field in our region of interest. To prove this conclusively, we explicitly estimate the BH losses on the OUV field.

Since we will conclude that energy losses on this target are not an important energy loss channel, the precise spectral shape is not a decisive feature for our purposes. We model the OUV field as a thermal distribution with a typical temperature $T_\mathrm{OUV}=3.33$~eV
\begin{equation}
    n_\mathrm{OUV}(E_\gamma)=\frac{15\phi L_X}{4\pi^5 r_\mathrm{OUV}^2T_\mathrm{OUV}^4}\frac{E_\gamma^2}{e^{E_\gamma/T_\mathrm{OUV}}-1},
\end{equation}
which is valid in the limit $r_\mathrm{OUV}\gg L$. We can now estimate the BH energy loss timescale off the OUV field as
\begin{equation}
    t_\mathrm{\mathrm{BH},\mathrm{OUV}}^{-1}(\gamma_p)=\frac{15\sigma_T\phi L_X m_e^4}{4\pi^5 r_\mathrm{OUV}^2 T_\mathrm{OUV}^4m_p}\frac{\xi_\mathrm{OUV}(\gamma_p)}{\gamma_p^2}.
\end{equation}
We can again determine the ratio between this energy loss timescale and the photohadronic timescale, noting that the function $\xi_\mathrm{OUV}(\gamma_p)/\gamma_p^3$ has a maximum value of about $9\times 10^{-25}$
\begin{equation}
    \frac{t_{p\gamma}}{t_\mathrm{BH,OUV}}\lesssim 3\times 10^{-3}\;\lcrg^2 M_7^{4/3} L_{X,43}^{-2/3}.
\end{equation}
We see that for the typical values considered in this work BH energy losses on the OUV photons are indeed negligible compared to the $p\gamma$ energy losses. The geometric reduction of the OUV signal inside the corona may raise tensions with the general explanation for the coronal X-rays as coming from the Comptonization of OUV photons. However, in principle X-rays could also originate from the Comptonization of the synchrotron photons from the accelerated electrons or of the low-energy photons injected in the photohadronic cascade that we have examined in the main text.

An additional source of cooling is synchrotron in the strong magnetic field. The typical energy-loss timescale is
\begin{eqnarray}
    &&t_\mathrm{syn}=\frac{3m_p^4c^3}{4\sigma_T u_B m_e^2 E_p}\\ \nonumber && \simeq 2.4\times 10^5\;\left(\frac{LM_7}{5r_g}\right)^2\frac{25\;\mathrm{TeV}}{E_p}L_{X,43}^{-1}\beta_{-1}\;\mathrm{s}.
\end{eqnarray}
which is again much longer that the $t_{p\gamma}$ loss timescale.

\section{Impact of fast cooling on the proton distribution}\label{app:strong_cooling}

In the main text, we have restricted ourselves to the regime where $E_{p,\mathrm{cool}}\gtrsim E_{p,\mathrm{br}}$, so that the protons carrying the most energy are not cooled significantly before escaping the layer. Here, we generalize our treatment to the fast-cooling case as well. For clarity, we denote by $\eta_p^*$ the fraction of magnetic energy density that would go into non-thermal protons in the absence of cooling. Thus, in the absence of cooling effects on spectral slopes, we would have
\begin{equation}
    \frac{dn_p^{\mathrm{(uncool)}}}{dE_p}=\frac{\eta_p^* u_B}{2E_{p,\mathrm{br}}^2}
    \begin{cases}        
    \left(\frac{E_p}{E_{p,\mathrm{br}}}\right)^{-1},  E_p<E_{p,\mathrm{br}}\\
    \left(\frac{E_p}{E_{p,\mathrm{br}}}\right)^{-s},  E_{p,\mathrm{br}}<E_p<E_{p,\mathrm{rad}}.
    \end{cases}    
\label{eq:npdiff_uncool}
\end{equation}
where $s\simeq 3$ as in the main text.
Notice that, as compared to Eqs.~\ref{eq:npdiff} and~\ref{eq:proton_number_density} that refer to the marginally fast-cooling case with $E_{p,\mathrm{cool}}\simeq E_{p,\mathrm{br}}$, the numerical factor is  different here since the spectrum after the break decreases as $E_p^{-3}$ rather than $E_p^{-4}$. Therefore, in the marginally fast-cooling case with $E_{p,\mathrm{cool}}\simeq E_{p,\mathrm{br}}$, the relation between the fraction $\eta_p$ we used there and the fraction $\eta_p^*$ defined here is $\eta_p=3\eta_p^*/4$.

Accounting for cooling, the steady-state spectrum can be approximated as
\begin{equation}
    \frac{dn_p}{dE_p}=\frac{dn_p^{\mathrm{(uncool)}}}{dE_p}\mathrm{min}\left[1,\frac{t_{p\gamma}}{t_\mathrm{esc}}\right].
\end{equation}
Therefore, the spectral shape of the protons depends on the hierarchy between $E_{p,\mathrm{br}}$ and $E_{p,\mathrm{cool}}$. The case $E_{p,\mathrm{cool}}\gtrsim E_{p,\mathrm{br}}$ was already discussed in the main text. In the case $E_{p,\mathrm{cool}}<E_{p,\mathrm{br}}$, protons cool significantly even below the break energy, leading to a typical spectrum
\begin{equation}
    \frac{dn_p}{dE_p}\propto
    \begin{cases}        
    \left(\frac{E_p}{E_{p,\mathrm{br}}}\right)^{-1},  E_p<E_{p,\mathrm{cool}}\\
    \left(\frac{E_p}{E_{p,\mathrm{br}}}\right)^{-2},  E_{p,\mathrm{cool}}<E_p<E_{p,\mathrm{br}}\\
    \left(\frac{E_{p,\mathrm{cool}}}{E_{p,\mathrm{br}}}\right)^{-2}\left(\frac{E_p}{E_{p,\mathrm{cool}}}\right)^{-s-1},  E_{p,\mathrm{cool}}<E_p<E^*_{p}\\
    \frac{E_{p,\mathrm{br}}^2 E_{p,\mathrm{cool}}^{s-1}}{(E_p^{*})^{s+1}}\left(\frac{E_{p}}{E^*_{p}}\right)^{-s},   E^*_{p}<E_p<E_{p,\mathrm{rad}}.
    \end{cases}    
\label{eq:npdiff_fastcool}
\end{equation}

\begin{figure}
    \centering
    \includegraphics[width=0.47\textwidth]{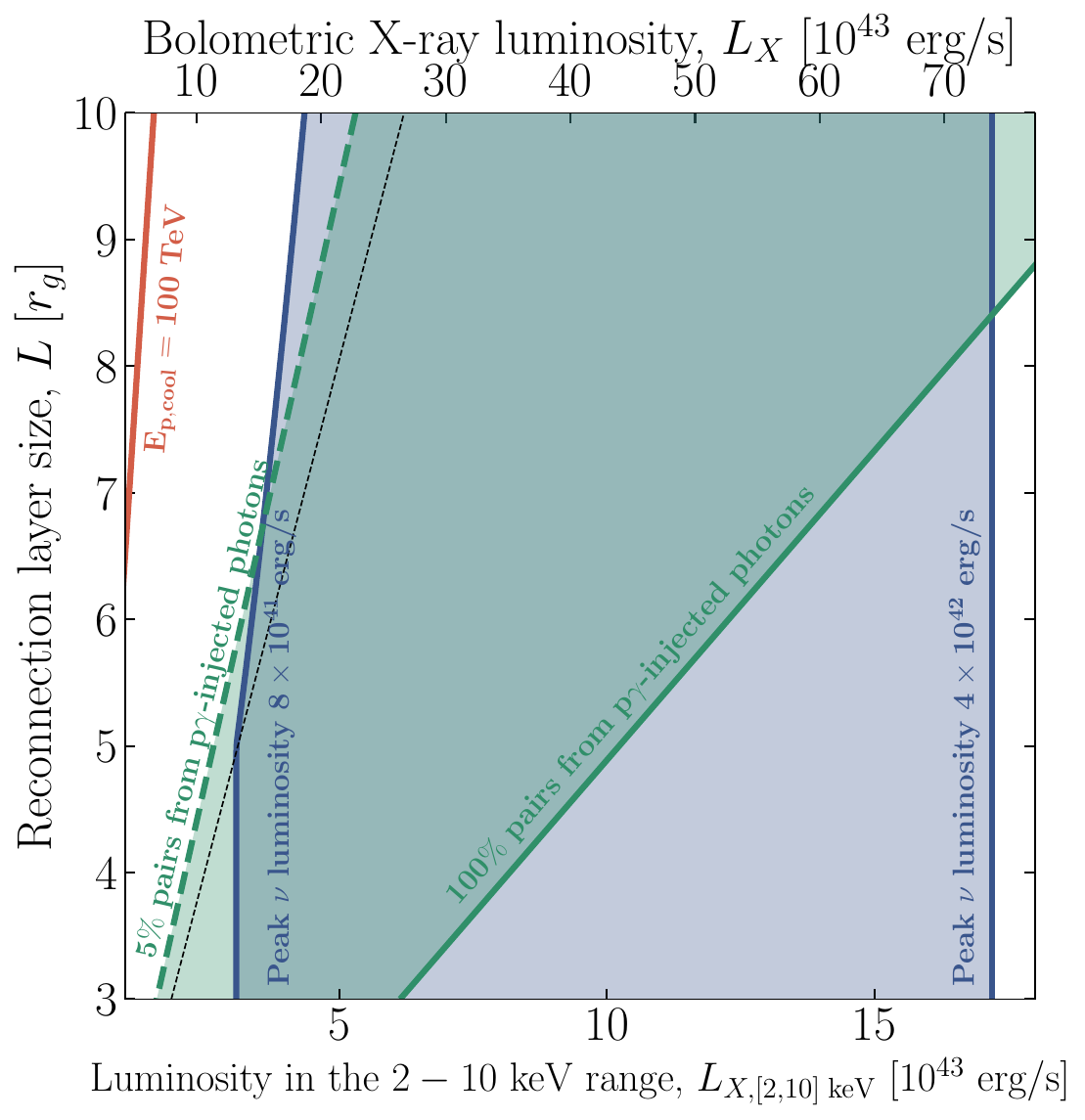}
    \caption{Same as Fig.~\ref{fig:example_regions}, accounting for the corrections arising in the fast-cooling regime. We choose $\eta_p^*=0.067$ to match the value of $\eta_p=0.05$ used in the marginally fast-cooling regime of Fig.~\ref{fig:example_regions}. Below the thin black dashed line, protons are fast-cooling ($E_{p,\mathrm{cool}}<E_{p,\mathrm{br}}$).}
    \label{fig:example_regions_cool}
\end{figure}

Therefore, the peak neutrino spectrum, which lies in the range $0.05E_{p,\mathrm{cool}}<E_p<0.05E_{p,\mathrm{br}}$, is
\begin{equation}
    (E_\nu^2 Q_\nu)^\mathrm{pk}=8.8\times 10^{40}\;\eta^*_{p,-1}L_{X,43}\mathrm{erg/s},
\end{equation}
namely, the neutrino luminosity is a fixed fraction of the X-ray luminosity, independently of the size of the reconnection region. This happens because the region becomes fully calorimetric, converting all of the proton energy via $p\gamma$ interactions.

The hadronically-produced pair number density would also be affected by the different form of the proton spectrum. Using the proton distribution accounting for cooling, we find
\begin{eqnarray}\label{eq:produced_pairs_cooling}
    \nonumber &&n_e^{\gamma\gamma}=  1.9\times10^6\;L_{X,43}^4 \eta_{p,-1}^{*2}\left(\frac{L M_7}{5 r_g}\right)^{-5}\left(\frac{E_{p,\mathrm{br}}}{25\;\mathrm{TeV}}\right)^2\;\mathrm{cm}^{-3}\\  && \mathrm{min}\left[1,\left(\frac{E_{p,\mathrm{cool}}}{E_{p,\mathrm{br}}}\right)^2\right].
\end{eqnarray}
In the fast-cooling regime $E_{p,\mathrm{cool}}<E_{p,\mathrm{br}}$, we obtain
\begin{equation}
    n_e^{\gamma\gamma}=7.2\times 10^8 L_{X,43}^2 \eta^{*2}_{p,-1}\left(\frac{L M_7}{5 r_g}\right)^{-3}\; \mathrm{cm}^{-3}.
\end{equation}

These results allow us to extend the regions in parameter space shown by Fig.~\ref{fig:example_regions} to the regime with $E_{p,\mathrm{cool}}\lesssim E_{p,\rm br}\simeq 25$~TeV. We show the corresponding regions in Fig.~\ref{fig:example_regions_cool}. For large X-ray luminosities, to the right of the thin black dashed line, we enter the fast-cooling regime, where the peak neutrino luminosity scales in proportion to the X-ray luminosity and becomes independent of the size of the reconnection region. This shows that the IceCube neutrino luminosity can also be explained in this fast-cooling regime, and that in fact an even larger amount of pairs can be attained in this regime compared to the marginally-fast cooling.

\section{Spherical geometry}\label{app:spherical_geometry}

The numerical code we use for the radiative calculations (see App.~\ref{app:numerical_approach}) is set up for a spherical geometry. Here we discuss the main changes induced by considering a spherical, rather than plane, region. To clearly differentiate between the two cases, we will define as $R$ the size of the spherical region. In relating the planar to spherical case, we will choose an effective size for the spherical region by equating the volumes for the sphere $4\pi R^3/3$ and the reconnection region $\beta_\mathrm{rec} L^3$. Therefore, $R$ is related to the full-length of the current sheet as $R = L (3\beta_\mathrm{rec}/(4 \pi))^{1/3}\simeq 0.28 \ L \beta_{-1}^{1/3}$.

The energy density of X-rays is now connected to the luminosity by

\begin{equation}
    u_X=\frac{3L_X}{4\pi R^2 c}.
\end{equation}
From the equality
\begin{equation}
   {F_X = \frac{L_X}{4 \pi R^2 }= 
    \eta_X \frac{c}{4\pi} \beta_{\rm rec}B^2}
\end{equation}
we obtain the magnetic energy density
\begin{eqnarray}
&&u_B =  \frac{ u_X }{6 \eta_X \beta_{\rm rec}}\\ \nonumber && \simeq 4.8\times 10^6 L_{X,43}\left(\frac{R M_7}{5r_g}\right)^{-2} {\beta_{-1}^{-1}} \, \rm erg\,cm^{-3}.
\end{eqnarray}

\exclude{
\begin{eqnarray}
    u_B&=&\frac{ u_X}{4\eta_X\beta_\mathrm{rec}}\simeq 2\times 10^6 L_{X,43}\left(\frac{R M_7}{5r_g}\right)^{-2} {\beta_{-1}^{-1}} \, \rm erg\,cm^{-3}.
\end{eqnarray}
}

The corresponding magnetic field is
\begin{equation}
    B\simeq 1.2\times 10^4 \left(\frac{R M_7}{5r_g}\right)^{-1} L_{X,43}^{1/2} \beta_{-1}^{-1/2}\;\mathrm{G}. 
\end{equation}
The acceleration timescale is also changed as
\begin{equation}
    t_\mathrm{acc}\simeq 2.5\times 10^{-3} \frac{E_p}{25\;\mathrm{TeV}}\frac{R M_7}{5r_g}L_{X,43}^{-1/2}\beta_{-1}^{-1/2}\;\mathrm{s}.
\end{equation}
The escape timescale is left unchanged and is
\begin{equation}
    t_\mathrm{esc}\simeq \frac{R}{c}\simeq 250 \;\frac{RM_7}{5r_g}\;\mathrm{s}.
\end{equation}
For the photohadronic timescale, the change in the X-rays energy density leads to the revised estimate
\begin{equation}
    t_{p\gamma}\simeq 2\times 10^4\;\left(\frac{R M_7}{5r_g}\right)^2 \frac{25\;\mathrm{TeV}}{E_p} L_{X,43}^{-1}\;\mathrm{s}.
\end{equation}
Equating this with the escape timescale we find the critical energy
\begin{equation}
    E_{p,\mathrm{cool}}=2.1\;L_{X,43}^{-1}\;\frac{RM_7}{5r_g}\;\mathrm{PeV}.
\end{equation}
We similarly find the maximum cooling-limited proton energy as
\begin{eqnarray}
    &&E_{p,\mathrm{rad}}\simeq 601\;\left(\frac{R M_7}{5 r_g L_{X,43}}\right)^{1/2}\beta_{-1}^{3/4}\\ \nonumber &&\mathrm{min}\left[1,5.9\left(\frac{R M_7}{5 r_g L_{X,43}}\right)^{1/2}\beta_{-1}^{-1/4}\right]\;\mathrm{PeV}.
\end{eqnarray}

In the peak neutrino spectrum, we have to account for an additional factor $(2\pi)\times (4\pi/3)$ in the denominator coming from the geometrical factors in $u_p\propto u_B$ and in the target photon density, and a correction factor $4\pi/3\beta_\mathrm{rec}$ to account for the relative volume factor between the spherical volume $4\pi R^3/3$ and the volume of the layer $L^3\beta_\mathrm{rec}$, leading to the estimate
\begin{equation}
    (E_\nu^2 Q_\nu)^\mathrm{pk}\simeq  10^{40}\;\frac{5r_gL_{X,43}^2\eta_{p,-1}}{RM_7\beta_{-1}}\frac{E_{p,\mathrm{br}}}{25\;\mathrm{TeV}}\;\mathrm{erg/s}.
    \label{eq:Lnu_pk_sphere}
\end{equation}
Finally, in the estimated number density of pairs, assuming that the fraction of MeV photons can be estimated just as in the planar case, we find
\begin{equation}
    n_e^{\gamma\gamma}\simeq 3.5\times 10^5\; L_{X,43}^4 \eta_{p,-1}^2\left(\frac{R M_7}{5 r_g}\right)^{-5}\left(\frac{E_{p,\mathrm{br}}}{25\;\mathrm{TeV}}\right)^2\beta_{-1}^{-2}\;\mathrm{cm}^{-3}.
    \label{eq:negg-sphere}
\end{equation}

Notice that in the spherical case, with $R\sim 0.28 L$ as discussed above, the pair number density can be much larger compared to the planar case (see Eq.~(\ref{eq:produced_pairs})), due to the spherical region being significantly smaller in all directions. Nonetheless, differences found in $n_e^{\gamma\gamma}$  between the planar and spherical geometry do not affect proton cooling, and in turn our neutrino estimates.

\begin{figure}
    \centering
    \includegraphics[width=0.47\textwidth]{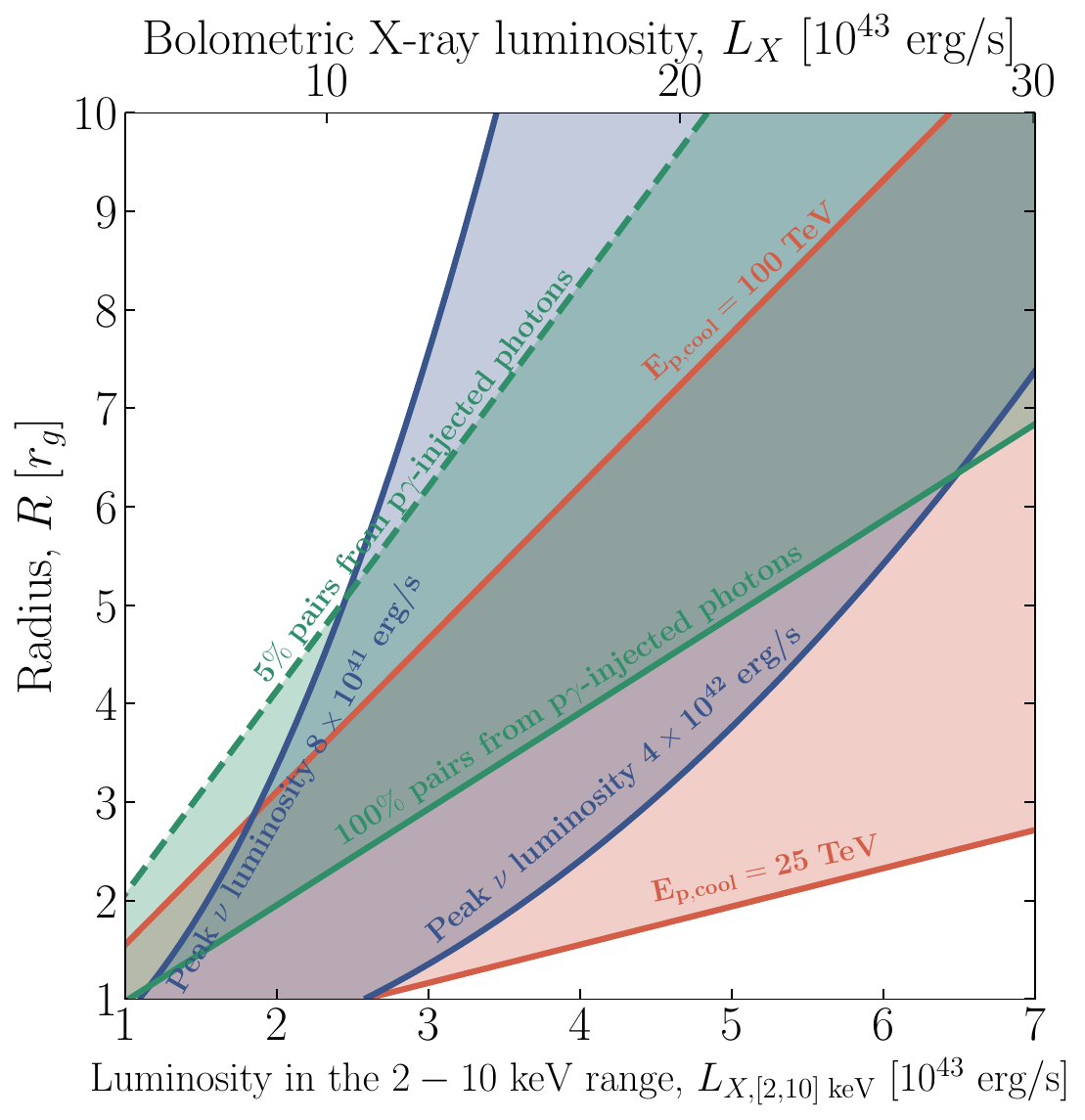}
    \caption{Same as Fig.~\ref{fig:example_regions}, but for spherical geometry. We show the regions in the parameter space $R-L_{X,[2-10]\;\mathrm{keV}}$, where $R$ is now the size of the spherical region.}
    \label{fig:example_regions_sphere}
\end{figure}

For completeness we show in Fig.~\ref{fig:example_regions_sphere} the regions of the $R-L_{X,[2-10]\;\mathrm{keV}}$ leading to a pair production comparable with the pair population inferred from Compton opacity, a peak neutrino flux consistent with the IceCube measurements, and  $25\;\mathrm{TeV}<E_{p,\mathrm{cool}}<100$~TeV. Compared to the planar case of Fig.~\ref{fig:example_regions}, for a compact enough region ($R\sim 1-2\;r_g$) one can simultaneously produce a sizable amount of pairs and saturate the observed neutrino flux, without entering the fast-cooling regime (see also Figs.~\ref{fig:sed} and \ref{fig:spectra}). This is due to the more compact spherical geometry, which allows for a larger production even at lower luminosities. 

\section{Numerical approach}\label{app:numerical_approach}

We supplement our analytical calculations of the neutrino spectrum with numerical results obtained with the proprietary code {\sc athe$\nu$a} \citep{2012A&A...546A.120D}. The code solves the kinetic equations for relativistic protons, secondary electrons and positrons, photons, neutrons, and neutrinos. The emitting region is assumed to be spherical with radius $R$. The magnetic field inside the source has strength $B$, and all particle populations inside the source are assumed to be isotropically distributed. The physical processes that are included in {\sc athe$\nu$a} and couple the various particle species are: electron and proton synchrotron emission, synchrotron self-absorption, electron inverse Compton scattering, $\gamma \gamma$ pair production, proton-photon (Bethe-Heitler) pair production, proton-photon and neutron-photon pion production. The photomeson interactions are modeled based on the results of the Monte Carlo event
generator {\sc sophia} \citep{2000CoPhC.124..290M}, while Bethe-Heitler production is modeled using the Monte Carlo results of \cite{1996APh.....4..253P}; see also \cite{2005A&A...433..765M}. For the modeling of other processes we refer the reader to \cite{1995A&A...295..613M} and \cite{2012A&A...546A.120D}. The code also returns the Thomson optical depth of the source based on the density of electron-positron pairs that cool due to radiative losses down to $\gamma_e = 1$. We note, however, that these cold pairs are ``passive'', because radiative processes of non-relativistic electrons (e.g., cyclotron radiation and Compton scattering) are not included in the numerical code. 
With the adopted numerical scheme, energy is conserved in a self-consistent way, since all the 
energy gained by one particle species has to come from an equal amount of energy lost by another particle species. The
adopted numerical scheme is ideal for studying the development of in-source electromagnetic cascades in both linear \citep[][]{2020ApJ...889..118Z} and non-linear regimes where the targets for photohadronic interactions are themselves produced by the proton population \citep[e.g.][]{2014MNRAS.444.2186P, 2021ApJ...906..131M}. 

As discussed in App.~\ref{app:spherical_geometry}, we consider a spherical region with a radius $R = L (3\beta_\mathrm{rec}/(4 \pi))^{1/3}\simeq 0.28 \ L \beta_{-1}^{1/3}$.
All particles escape from the source on an energy-independent timescale $t_\mathrm{esc}=R/c$. Relativistic protons, after being accelerated into a broken power-law distribution by reconnection (see also Sec.~\ref{sec:specprot}), are then ``injected'' into the spherical region at a rate given by
\begin{equation} 
\frac{{\rm d}q_p}{{\rm d}E_p} = q_0
\begin{cases}
\left( \frac{E_p}{E_{p, \rm br}}\right)^{-1}, \, E_p \le E_{p, \rm br}\\
\left( \frac{E_p}{E_{p, \rm br}} \right)^{-s}, \, E_{p, \rm br} < E_p \le E_{p, \rm rad}
\end{cases}
\end{equation}
where $q_0 =(s-2)/(s-1) \eta^{*}_{p} c u_B R^{-1}  E_{p, \rm br}^{-2} $, assuming $E_{p, \rm rad} \gg E_{p, \rm br} \gg E_{p, \min}$. Here, we introduce $\eta_p^*$ to refer to the fraction of energy carried by the proton population in the absence of cooling, as explained in App. \ref{app:strong_cooling}. To compute the steady-state proton distribution and the emerging photon and neutrino emission, we study the system for $10 R/c$; this is also a typical lifetime of the reconnection region, for $\beta_{\rm rec}\sim0.1$. Results are shown  for $M=6.7 \times 10^6 \ M_\odot$, $E_{X,\min}=100$~eV, $E_{X,\max}=100$~keV,  X-ray photon index -2, $L_X = 5\times 10^{43}$~erg s$^{-1}$, $\beta_\mathrm{rec}=0.1$, $\eta_X=0.5$, $\eta_p^*=0.1$, $s=3$, $E_{p,\min}=m_p c^2$, $E_{p, \rm br}=25$~TeV, $E_{p, \rm rad}=100$~PeV, $B=10^{5.1}$~G, and $R\simeq 1.4 r_g \simeq 1.4\cdot 10^{12}$~cm. 

\begin{figure}
\centering
    \includegraphics[width=0.45\textwidth]{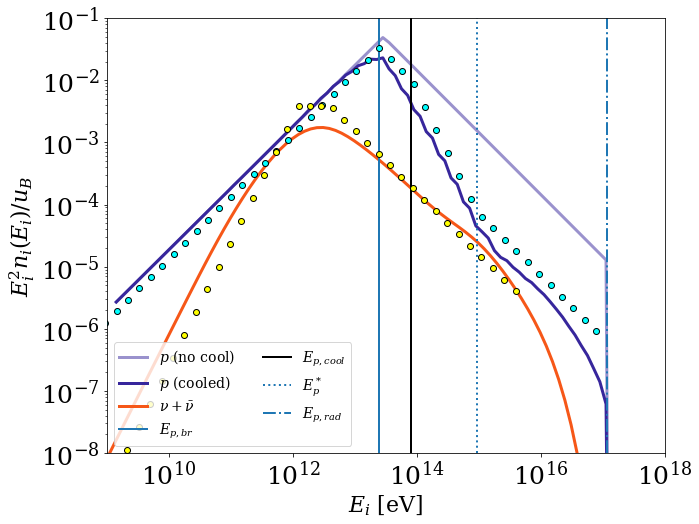}
    \caption{Proton (thick blue line) and all-flavor neutrino (thick orange line) energy density spectra normalized to the magnetic energy density. The steady state proton distribution has $\eta_p=0.05$. The spectrum of protons ``injected'' into the radiative region with $\eta_p^* = 0.1$ (i.e., the reconnection downstream) is overplotted with a transparent solid line. Vertical lines indicate characteristic proton energies discussed in the text. Protons with energies $E_p \gtrsim E_{p, \rm cool}$ cool due to photomeson interactions with the X-ray coronal photons. Cyan and yellow markers indicate the analytical expression for the steady-state proton and neutrino energy distribution, which agrees well with the numerical solution.}
    \label{fig:spectra}
\end{figure}

Figure~\ref{fig:spectra} shows the steady-state proton distribution we obtain numerically (thick solid line) and analytically as explained in Sec.~\ref{sec:specprot} (cyan markers). The analytical expression describes well the spectral break at $E_{p}^*$, caused by the change in the energy dependence of the photopion loss timescale. 

Some key findings from the numerical analysis:
\begin{itemize}
    \item Protons with energies $E_p \gtrsim E_{p, \rm cool}$ cool efficiently through photomeson interactions (equivalently, the reconnection region is optically thick to photomeson interactions for protons beyond this energy). 
    \item About 50\% of the energy carried initially by the proton population into the radiative zone (i.e., the reconnection downstream) is transferred to secondary particles (resulting in $\eta_p \simeq 0.05$), with neutrinos taking about 12\% of the energy lost by  protons.
    \item We find that the  density of pairs with $\gamma_e \sim 1$ is  $n_{e, \rm cold} \sim 3 \times 10^{11}$~cm$^{-3}$. This value is consistent with the analytical estimate in Eq.~\ref{eq:negg-sphere} for $\eta_p =0.05$ and $R=1.4 r_g$. Interestingly, it is also  close to the pair density needed to establish a Thomson optical depth of order unity in the corona (see Eq.~\ref{eq:ne}).
\end{itemize}

\section{Neutrino emission outside of the reconnection layer}\label{app:emission_outside}

Protons escaping the reconnection layer can provide an additional contribution to the neutrino emission. Photohadronic neutrino production is generally dominated by the reconnection layer, due to the rapid geometric dilution of both the emitted proton population and the X-ray photon field. On the other hand, a reasonable concern is whether proton-proton scattering on the material in the accretion flow can significantly contribute to the neutrino emission. We now show that this is not the case, by estimating the energy-loss timescale due to $pp$ scattering on the accretion flow gas.

The proton density in the accretion disk can be estimated as
\begin{equation}\label{eq:acc_number_density}
    n_p^\mathrm{acc}\simeq \frac{\dot{M}}{4\pi r H v_r m_p},
\end{equation}
where $H$ denotes the disk's scale height, and $v_r$ is the radial inflow velocity. Considering that the inner disk is thick, with $H\sim r$, due to the disk expansion associated with radiation pressure, and taking $v_\phi\sim(GM/r)^{1/2}$ and $v_r\sim\alpha v_\phi$ as the azimuthal and the radial velocity, with $\alpha\sim 0.1$ being the standard $\alpha$ parameter from Shakura and Sunyaev theory of accretion disks, we use Eq.~\ref{eq:mdot} and obtain
\begin{equation}
    n_p^\mathrm{acc}\simeq 7\times 10^{10}\frac{L_{X,43}\phi_1}{\alpha\, \eta_{\mathrm{acc},-1} M_7^2}\left(\frac{L}{5r_g}\right)^{-3/2}\;\mathrm{cm}^{-3}.
\end{equation}
where $\phi=10\,\phi_1$.
We can now estimate the $pp$ energy-loss timescale as (see, e.g.,~\cite{Murase:2017pfe})
\begin{eqnarray}
    &&t_{pp}=(n_p^\mathrm{acc}\sigma_{pp}\kappa_{pp}c)^{-1}\\ \nonumber &&\simeq 3\times 10^4\;\frac{\alpha\eta_{\mathrm{acc},-1}M_7^2}{L_{X,43}\phi_1}\left(\frac{L}{5r_g}\right)^{3/2}\;\mathrm{s},
\end{eqnarray}
where $\sigma_{pp}=3\times 10^{-26}$~cm$^2$ is the $pp$ cross section and $\kappa_{pp}=0.5$ is the inelasticity. In order to see whether $pp$ interactions can be fast enough to significantly produce neutrinos, we should compare this timescale with the typical residence time of the protons in the inner region of the accretion disk, after they escape the reconnection layer. If protons are able to freely stream out of the region, such timescale is of the same order of magnitude as the escape timescale from the layer, and therefore the $pp$ production is negligible. On the other hand, if proton escape is diffusive, their residence time in the disk would be larger, possibly leading to an enhanced $pp$ production. In any case, since only protons below the cooling energy $E_{p,\mathrm{cool}}$ are able to escape the reconnection layer while still retaining their energy, this additional channel of neutrino production would mostly be relevant at energies near or below the neutrino peak, and therefore would not be observable at IceCube.

A final possibility that one could consider is that neutrons, produced in the reconnection layer by $p\gamma$ interactions, could escape from it and decay outside, producing neutrinos. However, since the fraction of neutron energy carried by a neutrino in beta decay is typically of the order of $10^{-4}$ (see, e.g.,~\cite{Lipari:2007su,Fiorillo:2021hty}), these neutrinos would appear at much lower energies, in the GeV range. Therefore, we do not account for this contribution in our work.

\end{document}